\documentclass[twocolumn, times, tighten]{aastex60} 
\usepackage{CJK}
\usepackage{textcomp} 
\usepackage{amsmath,amstext}
\usepackage{newtxmath}
\usepackage{color} 
\usepackage{amssymb}
\usepackage{hyperref}
\graphicspath{{./}{figs/}} 
\def\GALEX{{\it GALEX}} 
\def\Swift{{\it Swift}}
\newcommand{\angstrom}{{\rm \mathring A}}

\newcommand{\hybeta}{\hbox{H$\beta$}}
\newcommand{\ciii}{\hbox{C\,{\sc iii]}}}
\newcommand{\mgii}{\hbox{Mg\,{\sc ii}}}
\newcommand{\feii}{\hbox{Fe\,{\sc ii}}}

\begin{document} 
\begin{CJK*}{UTF8}{gbsn} 
\revised{\bf Draft: \today} 
\title{On the UV/optical variation in NGC 5548: new evidence against the
reprocessing diagram}

\author{ Fei-Fan Zhu (朱飞凡)\altaffilmark{1,2,3}, Jun-Xian Wang
(王俊贤)\altaffilmark{1,2}, Zhen-Yi Cai (蔡振翼)\altaffilmark{1,2}, Yu-Han Sun
(孙玉涵)\altaffilmark{1,2}, Mou-Yuan Sun (孙谋远)\altaffilmark{1,2}, Ji-Xian
Zhang (张继贤)\altaffilmark{1,2}} 
\altaffiltext{1}{CAS Key Laboratory for Research in Galaxies and Cosmology,
    Department of Astronomy, University of Science and Technology of China,
    Hefei 230026, China} 
\altaffiltext{2}{School of Astronomy and Space Science, University of Science
    and Technology of China, Hefei 230026, China} 
\altaffiltext{3}{Department of Astronomy and Astrophysics, 537 Davey Lab, The
    Pennsylvania State University, University Park, PA 16802, USA}

\email{zff1991@mail.ustc.edu.cn, jxw@ustc.edu.cn, zcai@ustc.edu.cn}
\begin{abstract} The reprocessing scenario is widely adopted in literature to
    explain the observed tight inter-band correlation and short lags in the
    UV/optical variations of active galactic nuclei (AGNs).  In this work we
    look into the color variability of the famous Seyfert galaxy NGC 5548 with
    high quality \Swift{} multi-band UV/optical light curves.
We find the color variation of NGC 5548 is clearly timescale-dependent, in a way
    that it is more prominent at shorter timescales.
    This is similar to that previously detected in quasar samples, but for the
    first time in an individual AGN.  We show that while a reprocessing model
    with strict assumptions on the driving source and the disk size can
    apparently match the observed light curves and inter-band lags, it fails to
    reproduce the observed timescale dependency in the color variation.  Such
    discrepancy raises a severe challenge to, and can hardly be reconciled under
    the widely accepted reprocessing diagram. It also demonstrates that the
    timescale dependency of the color variation is uniquely powerful in probing
    the physics behind AGN UV/optical variations.  \end{abstract}

\section{Introduction} \label{sect:intro} It has long been known that the
ultraviolet (UV) and optical continuum radiation in active galactic nuclei
(AGNs) are variable \citep[e.g.,][]{1997ARA&A..35..445U}.  Such variation, as a
characteristic feature of AGNs, is useful in many aspects of AGN studies.  It
can be used to separate AGNs from stars/galaxies
\citep[e.g.,][]{2007AJ....134.2236S,2009A&A...497...81B,2010ApJ...714.1194S,2011ApJ...728...26M,2011AJ....141...93B,2012ApJ...760...51R,2014MNRAS.439..703G,2015ApJ...811...95P}.
It is also essential to the reverberation mapping of AGNs
\citep[e.g.,][]{1993PASP..105..247P,2014SSRv..183..253P}.  Thanks to the
advancement of observational techniques and equipment, the variation studies of
AGNs have increasingly attracted more and more attention, particularly in the
current age of time domain astronomy.  However, the origin of and the physics
behind the variation itself remains poorly understood.

A remarkable feature of AGN's UV/optical variability is that the light curves of
different bands are well coordinated, with wavelength-dependent time delays
\citep[long wavelength light curves lag those short wavelength ones;
][]{1991ApJ...371..541K,1994ApJ...425..609S,1997ARA&A..35..445U,1998ApJ...500..162C,1998ApJ...505..594N,1998PASP..110..660P,2000ApJ...535...58K,2001ApJ...561..146C}.
A reprocessing scenario, in which the accretion disk is illuminated by a central
variable source and thus produces variable reprocessed UV/optical emission
\citep{1988MNRAS.233..475G}, is commonly invoked in literature to explain such
coordinated variability and short time delays.
In this scenario, the variability origin should be traced back to the central
radiation source, which is suspected to be the presumed `corona'
\citep[e.g.,][]{1991ApJ...380L..51H,1995ApJ...455..623C}, and the delay
corresponds to the light travel time. Refer to Section~\ref{sect:simu} and
Section~\ref{sect:diss} for a detailed description and discussion on the
reprocessing diagram.

Another prominent feature of the variability is that the variation amplitude is
wavelength dependent in the sense that shorter bands vary more significantly
\citep{1997ARA&A..35..445U}.  AGNs thus generally appear bluer when they
brighten up.  Such ``bluer-when-brighter'' (BWB) trend has been confirmed in
numerous papers
\citep[e.g.,][]{2012ApJ...744..147S,2014ApJ...783..105R,2016ApJ...822...26G}.
Recently, it is found that such BWB trend is timescale-dependent for quasars
(luminous AGNs), in the way that shorter term variation is even bluer, as
revealed by SDSS observations in optical bands \citep{2014ApJ...792...54S} and
\GALEX{} data in ultraviolet bands \citep{2016ApJ...832...75Z}.  As demonstrated
with Monte-Carlo simulations, such timescale dependence of the BWB trend
potentially links the UV/optical variation in quasars to thermal fluctuations of
the accretions disk \citep{2016ApJ...826....7C}.  However, note that such
inhomogeneous disk model with completely independent thermal fluctuations is
unable to explain the inter-band lags generally observed in Seyfert galaxies.

It is known that in less luminous AGNs, such as Seyfert galaxies, X-ray
contributes to a larger fraction of bolometric luminosity
\citep{2010A&A...512A..34L,2005AJ....130..387S,2010ApJS..187...64G} and
reprocessing is expected to play a more dominant role.  Do Seyfert galaxies show
timescale-dependent color variation similar to their luminous analogs, i.e.,
quasars?  If yes, is such timescale-dependent color variability compatible with
the widely adopted reprocessing diagram? 

Recent multi-band monitoring campaigns of the well known Seyfert galaxy NGC 5548
have provided high quality light curves from far UV to optical
\citep{2015A&A...575A..22M,2016ApJ...821...56F}.  Historically, this source has
been studied over and over on its optical/UV variation and the lags within the
reprocessing diagram
\citep{1991ApJ...371..541K,2000ApJ...528..292C,2003ApJ...584L..53U,2006ApJ...639...46S,2007arXiv0711.1025G,2014MNRAS.444.1469M,2016ApJ...821...56F,2016MNRAS.456.1960S}.
In this work, to address the aforementioned questions,  we present a case study
using \Swift{} multi-band light curves of NGC 5548.   We give a brief
description of the photometric light curves gathered by \Swift{} in
Section~\ref{sect:data}.  In Section~\ref{sect:simu}, we construct a
reprocessing model to simulate multi-band light curves.  The artificial
multi-band light curves appear well matched to the observed ones, apparently
supporting the reprocessing diagram.  As shown in Section~\ref{sect:colorv}, NGC
5548 exhibits clear timescale dependence in its UV/optical color variation,
similar to quasars.  However, using simulated light curves we find that the
reprocessing diagram is unable to explain the observed timescale dependence,
thus is severely challenged.  Discussions and conclusions are given in
Section~\ref{sect:diss}.  Cosmological parameters $\Omega_m = 0.28$,
$\Omega_\Lambda = 0.72$, and $H_0 = 70\, \mathrm{km}$\,
$\mathrm{s}^{-1}\,\mathrm{Mpc}^{-1}$ \citep{2011ApJS..192...18K} are used where
necessary throughout this work.
 
\section{\Swift{} light curves} 
\label{sect:data} We glean \Swift{} light curves on NGC 5548 presented in
\citet{2015ApJ...806..129E} \footnote{\citet{2016ApJ...821...56F} also made use
of this set of data, and provided multi-band overlapping UV/optical light curves
obtained with {\it HST} and ground based telescopes. Those light curves however
span significantly shorter time intervals, containing no information of color
variation at long timescales. In this work we utilize the \Swift{} light curves
only.} and carry out color variability analysis based on them.  The \Swift{}
monitored on NGC 5548 in six ultraviolet and optical bands (UVW2 centered at
1928 \AA,  UVM2 at 2246 \AA, UVW1 at 2600 \AA, {\it U} at 3465 \AA, {\it B} at
4392 \AA, and {\it V} at 5468 \AA
\footnote{\url{http://www.swift.ac.uk/analysis/uvot/filters.php}}).  Though the
UVW2 light curve is available from Feb. 2012 to Aug. 2014 with cadences as short
as 1 to 3 days and only small gaps ($\sim$ 2 months), for the other 5 bands the
light curves span a shorter period between Mar. 2013 to Aug. 2014.  Such well
sampled multi-band light curves provide an excellent chance to examine whether
the color variation is timescale dependent in an individual Seyfert galaxy.

We notice that light curves of the same source have also been analyzed in
\citet{2016A&A...588A.139M} with a few more observations included (obtained
between the end of Dec. 2014 and early Feb. 2015), but the light curves have
been binned with bin size of 2 days, missing precise timing information and
heavily affecting the determination of time lags.  The source has also been
observed a few times by \Swift{} during 2005 and 2007, and it is found that the
obscuration property was significantly different back then with X-ray spectral
analysis \citep{2016A&A...588A.139M} and we leave them out in research of color
variability to avoid complication.  Another reason that we do not include the
few observations obtained in 2005, 2007 and 2015 is that there are large gaps
between these data sets and the rest major data sets. Including them can mainly
contribute to the measurement of color variation on very long timescales but not
on shorter timescales, thus may produce bias while comparing the color variation
at various timescales. Meanwhile the observations we include are well sampled
with small gap, so each observation can contribute to the analyses on all
timescales.

\section{The reprocessing model} \label{sect:simu} To study whether the
reprocessing diagram could produce a timescale-dependent color variation
pattern, we first construct a simple irradiated accretion disk model.  A
standard thin accretion disk is divided into rings with the ratio of the outer
to inner radius of each ring fixed.  The most inner radius $r_{\rm in}$ is
selected as the inner-most stable circular orbit for a Schwarzschild black hole,
that is, $r_{\rm in} = 6R_{\rm g}$, where $R_{\rm g} \equiv GM/{c^{2}}$ is the
gravitational radius and $M$ stands for the mass of the central black hole.
Each ring is further azimuthally and equally divided into zones with the same
radius but different azimuth.  Under the reprocessing scenario, the radiation at
radius $r$ is determined by both the local viscous heating of the accretion
disk, and the radiation received by the disk from the central illuminating
source.  Based on the {\it Stefan-Boltzmann Law}, $F = \sigma T^4$, the
temperature profile $T(t, r, \theta)$ of the disk could be expressed as
\citep{2007MNRAS.380..669C} \begin{equation} \begin{split} T^4(t, r, \theta) =&
    \dfrac{3GM\dot{M}}{8\pi \sigma r^3}\left( 1-\sqrt{\dfrac{r_{\rm in}}{r}}
    \right) + \\ &(1-A) \dfrac{h_* L_*(t-\tau(r, \theta, i))}{4 \pi \sigma
r_*^3}, \end{split} \end{equation} where $r$ is the radius, $\theta$ the azimuth
angle, $M$ the black hole mass, $\dot{M}$ the accretion rate, $A$ the disk
albedo, $h_*$ the vertical distance from the central variable source to the
disk, $r_*$ the distance from the central variable source to disk elements which
equals $\sqrt{h_*^2 + r^2}$, and $L_*$ the luminosity of the central variable
source, delayed by the light travel time $\tau(r, \theta, i)$ for different
regions, $(r, \theta)$, and inclination angle, $i$, between the line of sight
and the disk axis.
The lag $\tau$ is determined by the position of the disk element and the disk
inclination angle  with the form \begin{equation} c\tau(r, \theta, i) =
\sqrt{h_*^2 + r^2} + h_*\cos i - r \cos \theta \sin i.  \end{equation} We list
necessary parameters for this reprocessing scenario in Table~\ref{tab:pars}.
Primary parameters, such as the black hole mass and the bolometric luminosity,
are gleaned from past literature on NGC 5548.  \begin{table} \centering
    \caption{Parameter values for the standard reprocessing model.}
    \begin{tabular}{cccc} \hline\hline Parameters         &     &  Values
        & Reference                \\ \hline Black hole mass&$M$&  $3.2 \times
        10^{7} M_{\odot}$& (1)    \\ Bolometric luminosity& $L_{\rm bol}$&
        $2.82 \times 10^{44}~\rm{erg~s}^{-1}$  & (2) \\ Radiative efficiency&
        $\eta$&  0.083       &        \\ Inclination angel&$i$   &  $45^{\circ}$
        & (1)                  \\ Redshift& z  &  0.017           &
        \\ Illuminating height& $h_*$               &  10~$R_{\rm g}$
        &  \\ Illuminating luminosity& $L_*$ &  0.15~$L_{\rm bol}$         &  \\
        Disk albedo & $A$ & 0.5 & \\ \hline \end{tabular} \begin{flushleft}
    References: (1) -- \cite{2014MNRAS.445.3073P}; (2) --
    \cite{2016A&A...587A.129E}.  \end{flushleft} \label{tab:pars} \end{table}

Having calculated all the temperature of the disk regions, the disk spectral
    energy distribution (SED) is simply the sum of the blackbody radiation of
    the disk regions. With appropriate input radiating light curve, we are able
    to acquire the variable multiple wavelength light curves driven by the input
    one, by convolving the SED with transmission
    files\footnote{\url{https://heasarc.gsfc.nasa.gov/docs/heasarc/caldb/data/swift/uvota/index.html}}
    of \Swift{} filters. 

\begin{figure}[!t] \centering \includegraphics[width=0.49\textwidth]{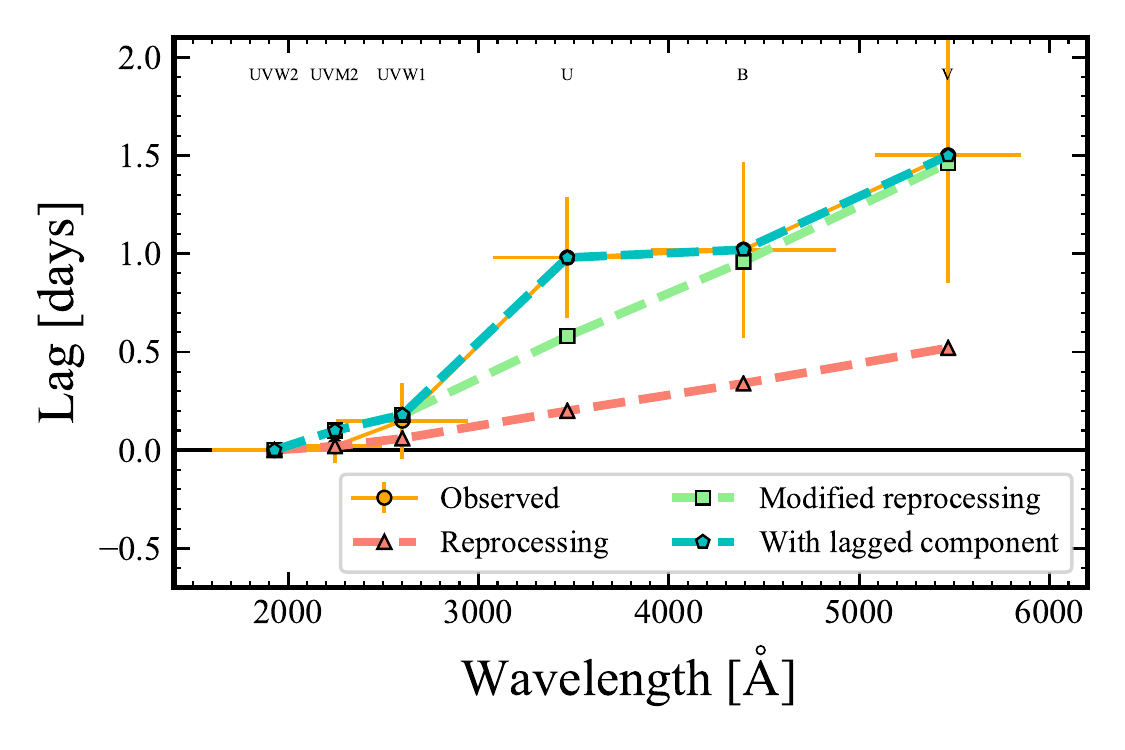}
    \caption{Observed inter-band lags in NGC 5548, comparing with our
    reprocessing models.  Lags are calculated between the six light curves and
    the one for UVW2, with the ICCF method detailed in
    \citet{2004ApJ...613..682P}. The observed lags are plotted as the orange
    solid line with error bars.  As for the simulations, we plot the lags
    predicted by both the original thin disk model (the light red dashed line)
    and the modified reprocessing model (the light green dashed line).
    Neglecting {\it U} band, the lags of the modified accretion disk model can
    broadly match the observed ones.  The model is further modified by
    considering an additional variable component, mimicking the effects of the
    variable but further lagged emission lines and Balmer continuum, for the
    three optical bands, {\it U}, {\it B}, {\it V}, so that the lags produced by
    the model can match observations.  The dashed dark cyan line which tracks
the observed lags almost perfectly represents the lags calculated for this
lagged component included model.} \label{fig:lagfit} \end{figure}

For NGC 5548, the illuminating source seems to be soft X-ray excess or FUV
    emission from the inner disk, rather than the X-ray corona \citep[][and see
    Section~\ref{sect:diss} for details]{2017MNRAS.470.3591G,
    2017ApJ...835...65S, 2017ApJ...840...41E}.  In this work, we do not
    associate the illuminating source with any wavelength at all since we do not
    have any valid information on it.  In the disk reprocessing model, only the
    variability pattern and luminosity of the light curve are needed.  We select
    the ultraviolet light curve collected at the shortest band, UVW2, which is
    also the most frequently sampled one, to model the variability pattern.  It
    is interpolated using the software package {\tt JAVELIN}
    \citep{2011ApJ...735...80Z,2013ApJ...765..106Z} to get better sampling.  As
    the geometry of the disk is already given, we can compute the resultant UVW2
    light curve based on the reprocessing model, as well as its time lag
    relative to the illuminating light curve.  The interpolated UVW2 light curve
    is then shifted backward in time by this lag\footnote{For a standard thin
    accretion disk with parameters listed in Table~\ref{tab:pars}, the lag is
    0.21 days. However, since the observed lags are a factor of 3 times larger
    than the value predicted by the standard thin disk model, we need to modify
    the disk model (see Figure~\ref{fig:lagfit}). The shift is computed to be
    0.61 days in this modified model.} to account for the expected lag between
    the illuminating source and the observed UVW2 band.  Starting from this
    illuminating light curve, the reprocessed light curve in each band and the
    predicted inter-band lags can then be derived.

In Figure~\ref{fig:lagfit} we plot the observed lags (peak values in ICCF)
    between other bands and UVW2. The expected values of the disk reprocessing
    models are over-plotted.  To calculate the inter-band lags, we adopt the
    traditional lag measurement method brought in \cite{2004ApJ...613..682P},
    interpolated cross-correlation function (ICCF), and implemented by
    \citep{Sun2018}\footnote{\url{http://ascl.net/code/v/1868}}.  The light
    curves have been detrended with a second-order polynomial linear
    least-squares fit separately before estimating the lags.

The inter-band lags in NGC 5548 were known systematically larger than predicted
    by the standard thin disk theory
    \citep{2014MNRAS.444.1469M,2015ApJ...806..129E,2016ApJ...821...56F}.  At
    large radii, within the standard reprocessing accretion disk model, lag
    $\tau$ measured for a given wavelength $\lambda$ follows a theoretical
    scaling relation of \citep{2016ApJ...821...56F} \begin{equation} \langle
        \tau(\lambda) \rangle \propto
    \left(\dfrac{k\lambda}{hc}\right)^{4/3}\left[\dfrac{3GM\dot{M}}{8\pi\sigma}
    + \dfrac{(1-A)L_*h_*}{4\pi\sigma}\right]^{1/3}.  \end{equation} Assuming
    that $(1-A)L_*H_*/r = \kappa GM\dot{M}/2r$, where $\kappa$ is the local
    ratio of external to internal heating \citep{2016ApJ...821...56F}, it can be
    simplified to \begin{equation} \langle \tau(\lambda) \rangle \propto
        \left(\dfrac{k\lambda}{hc}\right)^{4/3}\left[\dfrac{GM\dot{M}(\kappa +
        3)}{8\pi\sigma}\right]^{1/3}, \end{equation} namely, \begin{equation}
    \langle \tau(\lambda) \rangle \propto (M\dot{M})^{1/3}\lambda^{4/3}.
    \end{equation} As suggested in \citet{2016ApJ...821...56F}, the observed
    lags are a factor of 3 times larger than the prediction from standard thin
    disk theory with proper parameters list in Table~\ref{tab:pars}.  Here we
    simply adjust the $M\dot{M}$ of the model to produce lags matching the
    observed ones, and leave further discussion to Section~\ref{sect:diss}.
    Lags three times the prediction of standard thin-disk theory, and the above
    equation, would translate into $M\dot{M}$ a factor of 27 times larger.
Judging from Figure~\ref{fig:lagfit}, within the modified model
($M\dot{M}$ enlarged), the lags can better match the observed ones, but
deviation for {\it U} band is still quite obvious. This could be explained by
the contamination from other variable spectral components, including the blended
\feii{} emission and Balmer continuum
\citep[e.g.][]{2016ApJ...821...56F}\footnote{Although host galaxy and narrow
emission lines also contribute to the total emission, but it would not affect
the lag calculation as its flux remains stable for the timescales concerned in
this work.}. Also, the broad emission line \hybeta{} would contribute to the
fluxes of {\it B} and {\it V} band somewhat.  We take these effects into
consideration by including an extra variable and lagged component in each of the
three bands.  \footnote{Line contamination from \mgii{}, \feii{} and \ciii{} to
the three ultraviolet bands are neglected.  Even taken into consideration, in
all cases with different lags assumed (4 to 20 days, with contribution inferred
to be less than 2 percent), they do not show any noticeable effect on color
variation.} With the illuminating light curve serving as ionization source,
light curve of the emission line would mimic the variation behavior of the
illuminating light curve.  The illuminating light curve is first rescaled to
have the same mean flux level and standard deviation of the corresponding
reprocessed continuum light curve.  Then the light curves are shifted and
smoothed by 8 days, which is the approximate time delay between the \hybeta{}
emission line and the ionization continuum \citep[assumed to be $1367 \angstrom$
in ][]{2016ApJ...821...56F}.  This extra lagged and variable component is added
proportionally to the corresponding reprocessed continuum light curves to
construct the final simulated light curves for the three bands. To determine the
contribution proportion of the extra lagged and variable component, we adjust
the proportion until the lags calculated from the final light curves can match
observations.  The extra lagged and variable component is found to make up 12.5,
1.5 and 1.5 per cent for the {\it U}, {\it B} and {\it V} band, respectively.
As we can see from Figure~\ref{fig:lagfit}, the lag matching is done pretty well
now.  We note these fractions can not be directly compared with the line flux
contributions presented in \citet{2016ApJ...821...56F} which were obtained
through spectral decomposition.  This is because: (1) narrow lines are
non-variable; (2) we simply assume the lagged variable component has the same
variation amplitude (before smoothing) as that of the disk continuum; A larger
fraction is needed if the lagged component has smaller variation amplitude.
Nevertheless, with this approach, the contamination from broad emission lines
and Balmer continuum to the broad band light curves are effectively simulated.
We shall explain in the next section that, as we carry out the analysis based on
flux differences, these non-variable components, such as narrow emission lines
and host galaxy emission, do not affect the results in this work.

\begin{figure*}[!t] \centering \includegraphics[width=1.\textwidth]{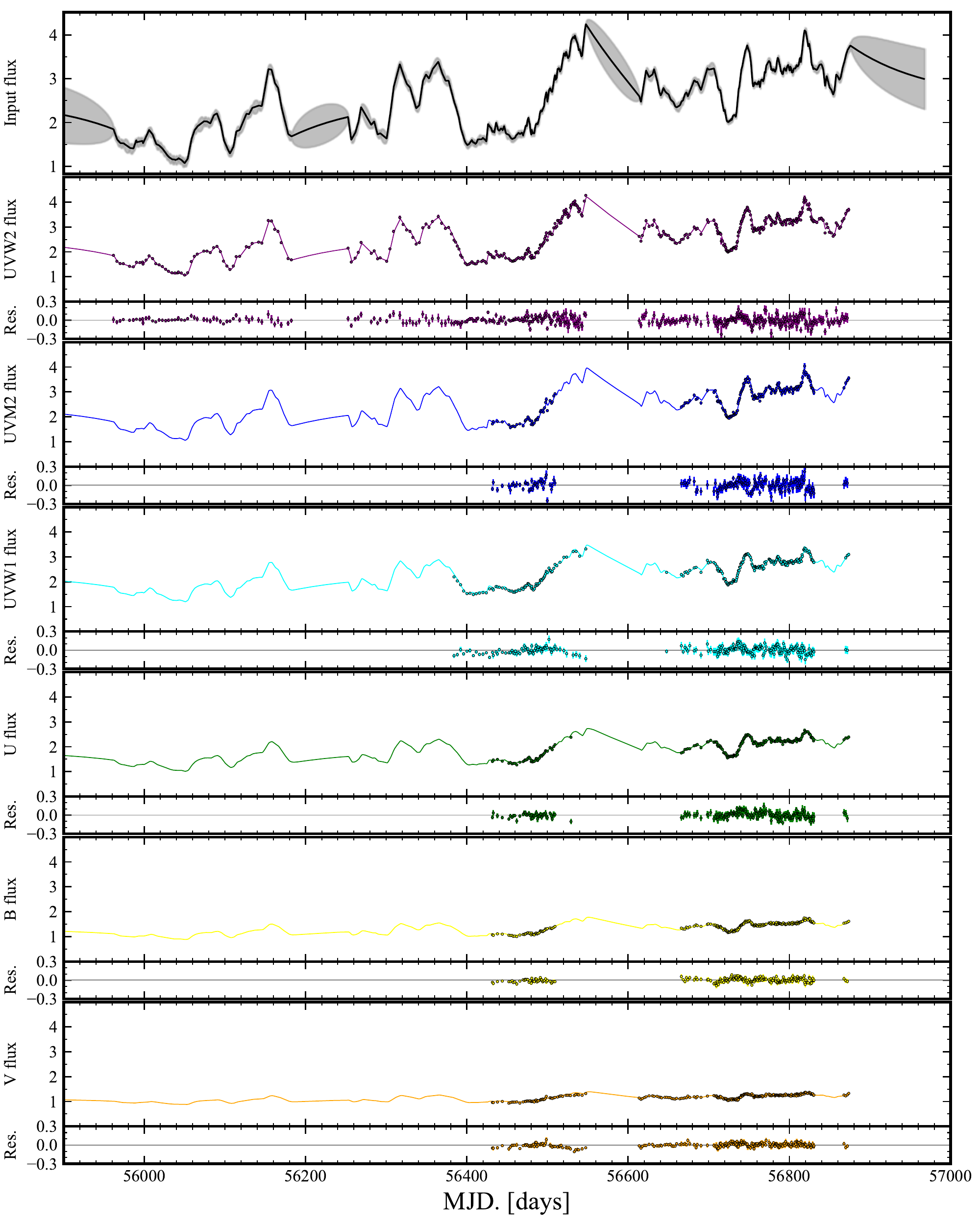}
    \caption{The observed light curves in unit of $10^{-14}~{\rm erg}~{\rm
    cm}^{-2}~{\rm s}^{-1} \angstrom^{-1}$ compared with those simulated ones
    from our reprocessing model. The extra lagged and variable component has
    been taken into consideration for the simulated ones. The top panel plots
    the light curve of the radiating source, generated using the software
    package {\tt JAVELIN}. In the following six panels, we plot both the
    observed and the simulated light curves of the six \Swift{} bands from the
    shortest wavelength to the longest. Observed fluxes are marked with filled
    dot and the simulated light curves are plotted as solid curves. The
    simulated light curves have been rescaled and shifted to match the variation
    amplitude and mean flux level of the observed ones. With this step, the
    baseline flux of the host galaxy is accounted for. At the lower part of each
panel, the difference between the observed flux and corresponding simulated flux
are plotted.} \label{fig:simlcs} \end{figure*}

The resultant simulated light curves of the six bands are still not
ready to be compared with the observed ones yet.  Using the observed light
curves as reference, the simulated light curves are shifted and rescaled.  The
shifting is to account for those non-variable component, such as radiation form
the host galaxy and slow-varying narrow emission lines.  It will not pose any
effect on the flux differences we used in the next section.  Rescaling is also
necessary.  As we do not know the exact wavelength range responsible for the
irradiation, we cannot constrain the variability amplitude of the illuminating
light curve.  As a direct result, variability of the emergent disk radiation of
the six bands is also unconstrained.  Also, the extinction effect of dust on the
observed fluxes are accounted for with the practice of rescaling.  Since it is
merely a linear transformation on the simulated light curves, rescaling is not
timescale dependent, and won't affect our timescale dependency analysis in the
next section. We plot the observed light curves together with the final
simulated light curves to make a direct comparison in Figure~\ref{fig:simlcs}.
Two sets of light curves match very well apparently.  That is to say, as already
demonstrated in previous study \citep{2017MNRAS.470.3591G},  the reprocessing
model with fine-tuned parameters seems to be able to well reproduce the observed
UV/optical variations in NGC 5548. 

\section{Timescale-dependent color variability} \label{sect:colorv}
Generally, color variability can be characterized using the ratio of the
flux/magnitude change of two bands \citep{2012ApJ...744..147S}. 
In this work, we follow a similar method. The ratio for any two epochs
on light curves of two bands is calculated as:

\begin{equation} s(\tau) = \frac{f^{\rm r}(t+\tau)-f^{\rm r}(t)}{f^{\rm
b}(t+\tau)-f^{\rm b}(t)}, \end{equation}

where $f^{\rm r}(t)$ and $f^{\rm b}(t)$ represent the flux measurements
in a redder and a bluer band at epoch $t$, respectively.  Only ratios with
$s(\tau) > 0$ are kept.  In addition, pair epochs in which the variation is
statistically insignificant ($< 3\sigma$) are excluded to avoid possible
unrealistic variations contributed by measurement uncertainties (please refer to
\citet{2014ApJ...792...54S} for details; the effect of measurement
uncertainties can also been seen in Section~\ref{sect:coverror}).  For those
$\tau$ falling into a certain timescale bin, the median value of these $s(\tau)$
are taken to mark the color variation amplitude of this timescale. In this way,
we can check for the timescale dependency of the color variability.

Ratio calculation done in flux space has the merit of avoiding host
galaxy contamination (host galaxy flux would cancel out), which is significant
for nearby AGNs, NGC 5548 for example.  On the other hand, extinction of
intervening dust would instead affect the flux differences.  But, as the dust
component would remain constant on timescales involved here (less than a year),
flux difference would be affected in the same manner for different timescales,
and the trend of the $s-\tau$ curve, which is really what we care about in the
following comparison, would remain unaffected. 

\begin{figure*}[!t] \centering
    \includegraphics[width=1.0\textwidth]{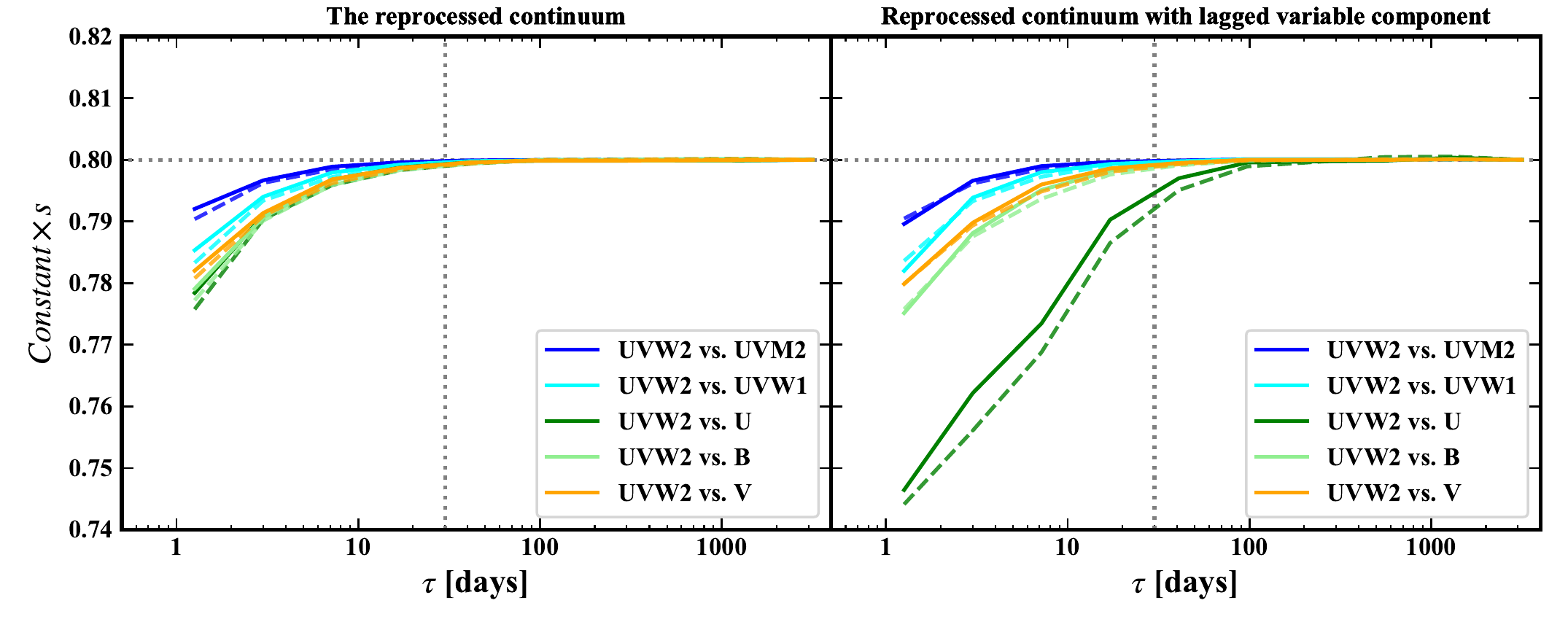} \caption{ The five lines
    of different colors represents the color variability results between UVW2
    and the other five different bands, namely, UVM2, UVW1, $U$, $B$ and $V$.
    These curves are multiplied by a constant to compare with the perfectly flat
    gray dotted line at $s = 0.8$. For both panel, the solid lines are the
    results under an illuminating light curve generated by a DRW model, while
    the dashed, lighter-colored lines are results calculated for a illuminating
    light curves with a PSD deviating from the DRW model.  {\it Left:} the
    dependency of color variability on timescale under the reprocessing
    paradigm, without considering the contribution from emission lines.  Under
    the reprocessing paradigm, the flat trend of these curves (at timescales
    above 10 days or so) indicates no timescale dependency, despite the
    different illuminating light curves.  {\it Right:} the dependency of color
    variability on timescale for the simulated light curves, with the lagged
    variable component taken into consideration.  The $s-\tau$ curves, except
    for the ones involving $U$ band, still remain quite flat on timescales over
30 days.  } \label{fig:simtheta} \end{figure*}

\subsection{Color variability under the reprocessing paradigm}
\label{sect:theory} Before set to handle the observed light curves, we first
check for the color variability under the simple reprocessing paradigm.  To
demonstrate if the reprocessing model can produce timescale-dependent color
variation in principle, here we generate artificial illuminating radiation light
curve lasting 5000 days based on the damped random walk (DRW) model
\citep{2009ApJ...698..895K}, with DRW parameters determined from modeling the
UVW2 (see Section~\ref{sect:simu} and Section~\ref{sect:diss} for why we choose
UVW2 to model the illuminating light curve) light curves with {\tt JAVELIN}
\citep{2011ApJ...735...80Z,2013ApJ...765..106Z}.  For the DRW model we adopted,
the deconvolution timescale, $\tau$, is 94.8 days, and the variability
amplitude, $\sigma$, is $0.743 \times 10^{-14}~{\rm erg}~{\rm cm}^{-2}~{\rm
s}^{-1} \angstrom^{-1}$ with a mean flux of $2.527 \times 10^{-14}~{\rm
erg}~{\rm cm}^{-2}~{\rm s}^{-1} \angstrom^{-1}$.  Note that using different DRW
parameters does not alter the results in this work.  With the illuminating light
curves, we run the reprocessing model, generate artificial light curves in the
six \Swift{} bands, and analyze the color variation behavior at different
timescales using these light curves.  For a simple illustration, we only carry
out the color variability calculation between the bluest band UVW2 and the rest
five redder bands (UVM2, UVW1, $U$, $B$, $V$).  Our results are presented as
solid curves in the left panel of Figure~\ref{fig:simtheta}.

At very short timescales (less than 10 days), the simulated light curves show
some timescale dependence in the color variation (the rising trend of the
$s-\tau$ curve).  And the rising trend is more prominent for two bands with
larger wavelength difference.  Clearly this short term rising trend is related
to the inter-band time lags and smoother variation at longer wavelengths.  Such
effects however are negligible on significantly larger timescales.  On
timescales larger than about 10 days, the $s-\tau$ curves appears quite flat,
indicating no timescale dependency at all, which contradict the
timescale-dependent color variability revealed with both SDSS and \GALEX{}
quasar samples \citep{2014ApJ...792...54S,2016ApJ...832...75Z}.

It should be noted that the choice of the model generating the input
illuminating light curve will not have any effect on this conclusion.  The
timescale independency is barely related to the PSD shape of the illuminating
light curve.  To better demonstrate this point, we test another illuminating
light curve, generated based on the method brought up in
\citet{1995A&A...300..707T}, and from the power spectral density (PSD) suggested
by \citet{2017arXiv170905271G}.  This PSD generally resembles the PSD of DRW
model, but has low frequency (less than about $10^{-6} Hz$) slope $-1.3$,
deviating from the flat slope of the DRW model at the same frequency range.  The
color variability results are presented as dashed, lighter-colored lines in
Figure~\ref{fig:simtheta}.  Their main features are the same as those solid
lines, despite the different PSDs of the illuminating light curves.

Also, to demonstrate the effect of emission lines on the timescale
independency of color variability, fluxes of them are added as detailed in
Section~\ref{sect:simu} for the three optical bands, $U$, $B$, $V$.  These
results are plotted in the right panel of Figure~\ref{fig:simtheta}.  Still, on
timescales longer than 30 days, the flat color variability curves remain mostly
unchanged, except for the $U$ band, whose $s-\tau$ curve shows a mild rising
trend on timescales from 30 days to 100 days.  This indicates that emission
lines do affect the timescale dependency of color variability, when the emission
lines are strong enough. For the other two bands suffering the same issue, the
timescale independency of color variability, possessed intrinsically by the
reprocessing model, remains untouched. 

\subsection{Color variability of NGC 5548} \label{sect:cv5548} Bearing the above
results in mind, we proceed to check for the observed color variability for NGC
5548 and compare with the reprocessing model. 

We carry out the color variability calculation for the two sets of light curves
obtained in Figure~\ref{fig:simlcs}: the observed one and the scaled simulated
one.  The simulated light curves have been scaled to account for the baseline
flux of the host galaxy.

\begin{figure*}[!t] \centering
    \includegraphics[width=1.\textwidth]{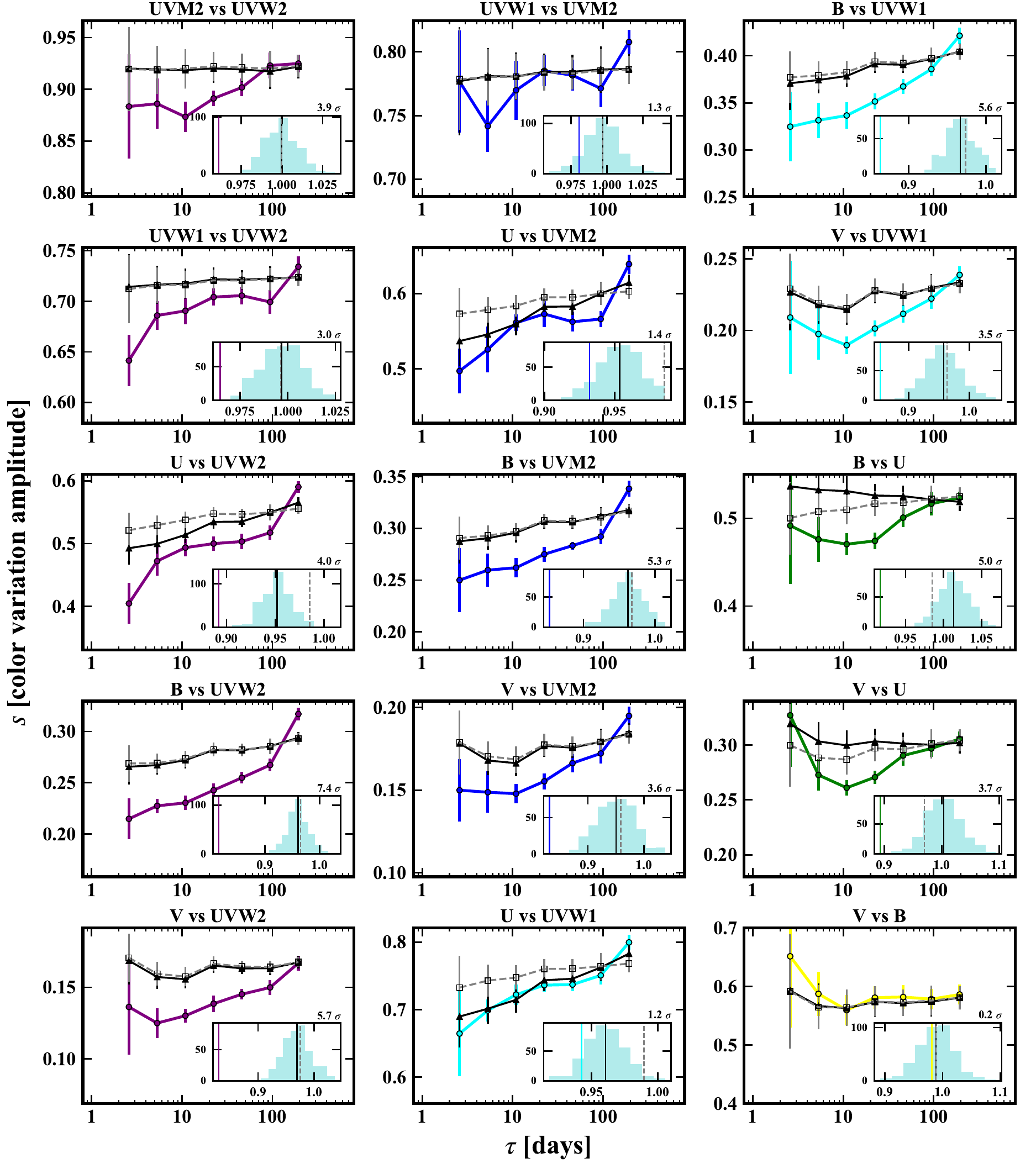} \caption{The timescale
    dependency of color variability between each pair of \Swift{} bands (marked
    above each frame).  In each panel, the colored solid line marked with filled
    circle represents the $s(\tau)$ calculated for the observed light curves.
    Color variability of the original simulated, emission line free and
    resampled light curves are plotted as the gray dashed lines in each panel.
    With proper lagged line component considered, new light curves with
    inter-band lags matching the observed ones are acquired, and their color
    variability is shown as the solid black curves.  In almost all panels, the
    observed $s(\tau)$ curve show a rising trend with increasing timescale
    compared with the simulated one on timescales from 10 days to 800 days.  To
    better illustrate this point, we add a sub-plot to compare observed and
    simulated $s(\tau)$ ratios for each panel ($s(4 < \tau < 34 days)/s(\tau >
    34 days)$; colored vertical line: observation; black: simulation).  The
    numbers above the upper right corner of these sub-plots demonstrate how
    significant the simulations deviate from the observations by
    measuring the separation of the two ratios with the standard deviation of
    the plotted distribution.  For the emission line free simulations, the ratio
    is plotted as the dashed gray vertical lines in the same sub-plots.
    Clearly, the simulated light curves cannot reproduce the rising trend seen
    in the $s-\tau$ curves of observed light curves.} \label{fig:alltt}
\end{figure*}

During the calculation, as the observed light curves are not observed strictly
simultaneously, the light curves need to be binned with bin size of 0.5 days
and, the photometric measurements of two bands are deemed to be observed
simultaneously if their observed time difference is less than 0.3 day.  To
estimate the uncertainties of the observed $s(\tau)$, we adopt a bootstrap
method resembling the one used by \cite{2004ApJ...613..682P} during the ICCF
practice for lags measurement.  In each bootstrap realization, $N$ random
selections are drawn, with replacement (so that the repeatedly selected
measurement would have some real effect on the outcome of the realization), from
the original $N$ photometric measurements in the light curve pairs.  New
$s(\tau)$ can be calculated from this new pair of light curves.  By repeating
such realization for 1000 times, we are able to build up a distribution for the
observed $s(\tau)$, and we take the standard deviation to be the measurement
uncertainties of $s(\tau)$.  In each sets, as there are six light curves and any
two of them can be used to generate a $s-\tau$ curve.  We result in a figure
with 15 frames as shown in Figure~\ref{fig:alltt}. 

In each frame, the colored curve represents the $s(\tau)$s calculated using real
\Swift{} light curves. 
The $s-\tau$ curves are clearly timescale-dependent with the majority of the
frames showing a rising trend, similar to those detected in the two
aforementioned quasar samples.  Note this is the first time that a
timescale-dependent color variation is reported in an individual AGN.  The fact
that the color variability of AGNs show clear timescale dependence can be used
to refute certain quasar variability mechanism, as demonstrated in
\cite{2014ApJ...792...54S}.  It also lends support to the inhomogeneous
accretion disk model brought up in \cite{2011ApJ...727L..24D} and revised in
\cite{2016ApJ...826....7C}.

In the last few frames of Figure~\ref{fig:alltt}, the observed
$s-\tau$ curve displays an opposite trend ($s(\tau)$ slightly decreases with
timescale), on the shortest timescales.  This signal is mainly connected to the
under-estimated photometric uncertainties.  On short timescales, variability
amplitude is much weaker, especially for these long wavelength bands.
Variability signal can be overwhelmed by their measurement uncertainties,
leaving the color variability signal noise-dominated.  We shall further expand
on this point in Section~\ref{sect:coverror}.

In addition, for those light curve pairs of neighboring filters, such as
UVW1 vs. UVM2, and $B$ vs. $V$, the rising trend is not so conspicuous.  This is
likely because the wavelength difference of the two bands is too small to
exhibit their different variability behavior.  For the rest band pairs that we
are concerned, the rising trend is {\it systematically} prominent.  We shall
then compare them with the $s-\tau$ curve obtained for the simulated light
curves.

Although we have already made the conclusion in Section~\ref{sect:theory} that
the reprocessing scenario in principle cannot reproduce the observed timescale
dependency, observational effects have not been taken into consideration.  In
view of the possible effect introduced by uneven sampling and photometric
uncertainties, we resample the simulated light curves presented in
Figure~\ref{fig:simlcs} based on the observed ones and add random flux
fluctuations based on observed photometric uncertainties.   Photometric
uncertainties are taken into consideration using the following equations:
\begin{equation} \label{eq:error} \begin{split} flux&(sim. + obs.) = flux(sim.)
+ \sigma, \\ \sigma &\sim Normal(0, \sigma(obs.)), \end{split} \end{equation}
where $sim.$ and $obs.$ stand for simulation and observation, respectively, and
$\sigma(obs.)$ represents the observed flux uncertainties. We carry out color
variability analysis on these resampled light curves. Such resampling procedure
are repeated for 400 times so that we can acquire the distribution of the
simulated $s(\tau)$ for all the timescale bin concerned.  So the observationally
affected simulated $s(\tau)$ and their associated uncertainties are chosen to be
the mean and variance of the distribution respectively.  The color variability
calculation for the simulated light curves also incorporates the 3-$\sigma$
criterion.  Also, the simulated light curves have been scaled to the observed
ones to account for the host galaxy contamination.  By doing so, we may then
make a more direct comparison of the observed $s-\tau$ relations with those
predicted by the reprocessing model.  The resulting $s-\tau$ curves based on the
emission line included simulated light curves are presented in each frame of
Figure~\ref{fig:alltt} as the solid black lines.  We also plot the results based
on the emission line free simulated light curves as the dashed gray lines for
reference.

To quantitatively compare the steepness/flatness of the $s-\tau$ curves, we
first average the $s(\tau)$ on both relatively shorter (between 4 and 34 days)
and longer (above 34 days) timescales.  The averaging are done on binned
$s(\tau)$ points in Figure~\ref{fig:alltt} rather than $s(\tau)$'s distribution
to avoid giving too much preference to those bins with more data pairs.  Then we
compute the ratios of the averaged $s(\tau)$ on shorter timescales to those on
relatively longer timescales.  So, a ratio value less than unity would indicate
a timescale dependency for the color variability, in which the color variability
is less conspicuous on longer timescales in comparison with that on shorter
timescales.  The smaller the ratio is, the more evident the timescale dependency
will be.  For the simulated $s-\tau$ curve, we are able to build up the
distributions for the ratios, based on the 400 times resampling of the dense
simulated light curves.  The difference between the observed ratios and the mean
values of the simulated ratios can be measured in unit of the variance of the
distribution.  In this way, we are able to tell if the difference is significant
enough between observed and simulated $s-\tau$ curves, and the distributions of
the ratios are also plotted in Figure~\ref{fig:alltt}, with the differences in
ratios marked on the upper right corner in the sub-plots.  By the way, we also
carry out the color variability for the emission line free light curves and the
results are plotted as the gray dashed lines in each frame.

As shown with the black solid curves in Figure~\ref{fig:alltt}, on timescales
above 4 days or so, the $s-\tau$ relations produced under the reprocessing
scenario, with observational effects taken into consideration, appear {\it
systematically} flatter than the observed ones in almost all the 15 frames.
This phenomenological conclusion is further buttressed by the ratio differences
we calculated above.  For the majority of the light curve pairs, the differences
are obvious, except for those band pairs involving $U$ band.  This indicates
that the contamination of the lagged emission lines on $U$ band can severely
impede the interpretation of color variability on short timescales, as we point
out in Figure~\ref{fig:simtheta}. 

\subsection{The effect of covariance error on color variability}
\label{sect:coverror} 
In this section, we further explore the effect of several measurement related
effect on the calculation of color variability.

\begin{figure*}[!t] \centering
    \includegraphics[width=1.\textwidth]{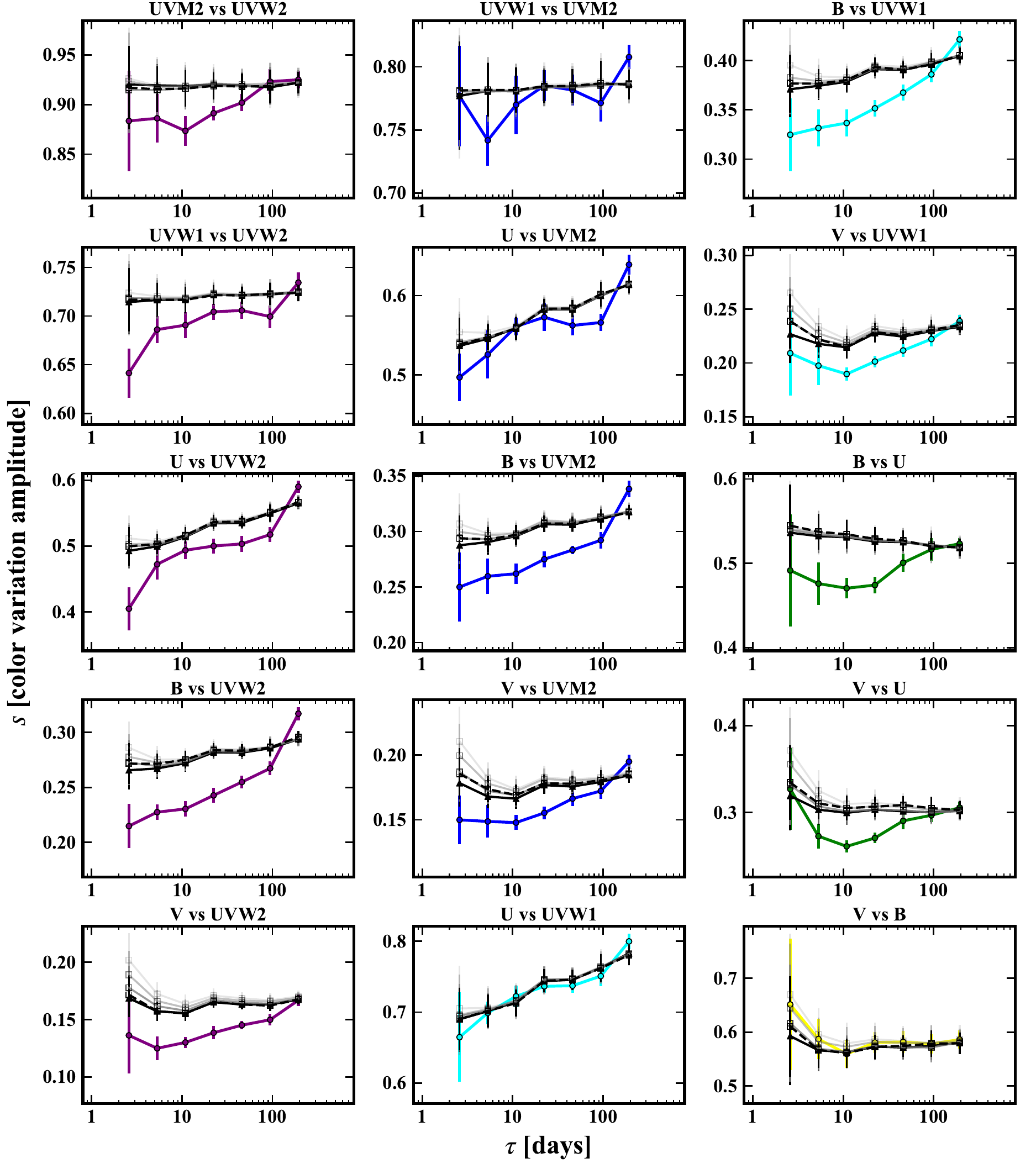} \caption{The
    effect of measurement uncertainties on the timescale dependency of color
    variability.  The 15 panels present results calculated for the 15 different
    pairs of bands.  In each panel, the solid colored curve is the observed
    color variability and the solid black curve represents result calculated for
    the simulated light curves.  The gray curves tell us what would happen to
    the $s-\tau$ curves if measurement uncertainties for the simulated light
    curves are underestimated.  And the dashed black curves show the impact of
    covariant uncertainties on the $s-\tau$ relations.  Under both scenarios,
    the $s-\tau$ curves would shift a bit upward, showing a mild declining trend
    on short timescales.  This indicates that neither one can be responsible
for the rising trend of the observed colored $s-\tau$ curves.}
\label{fig:coverror} \end{figure*}

We first look at the possible underestimation of \Swift{} observational
uncertainties.  We make use of photometric uncertainties in our simulation to
take observational effect into consideration.  If underestimated, uncertainties
would directly affect the color variability of simulated light curves.  This
effect is assessed by tweaking Equation~\ref{eq:error} just a little bit:
\begin{equation} \begin{split} flux&(sim. + obs.) = flux(sim.) + \sigma, \\
\sigma &\sim Normal(0, k \times \sigma(obs.)), \end{split} \end{equation} where
k is chosen to be 1.1, 1.2 and 1.3 in our three trials.  The color variability
results for these light curves are plotted in Figure~\ref{fig:coverror} as the
gray curves, with lighter gray curves representing results calculated using
larger k values.  It can be seen that, compared with the black curves, all the
gray curves show a mild declining trend on short timescales (contrary to
observations), with larger k value yielding more prominent upward tilt toward
shorter timescales.  Therefore underestimated photometric uncertainties (if
exist) would even strengthen our conclusions that the reprocessing diagram is
unable to reproduce the observed $s-\tau$ curves.

The possible measurement covariance between any two \Swift{} band photometric
observations is another issue.  It tends to introduce in-phase deviations to the
photometric measurements of two bands at the same time, hence affecting the
result of color variability.  As it is rather intricate to access the level of
the systematic covariance, here we check for their impact on $s-\tau$ curves by
adding them to the simulated light curves: \begin{equation} \begin{split}
    {flux}_{\rm band1} &= {flux}_{\rm band1} + k * \sigma_{\rm band1},  \\
{flux}_{\rm band2} &= {flux}_{\rm band2} + k * \sigma_{\rm band2},  \\ k &\sim
Normal(0, \sigma_{norm}), \end{split} \end{equation} where $\sigma_{norm}$ is
used to indicate the prominence of covariance uncertainties. We adopt 0.3 here
as a tryout.  The $s(\tau)$ calculation is done the same way for these light
curves, and the results are shown as the dashed black curves in
Figure~\ref{fig:coverror}.

By comparing the dashed black curves and the solid black ones, it can be seen
that when considering covariance uncertainties, the curves would also show a
mild declining trend on short timescale.  Different values for $\sigma_{norm}$,
ranging from 0.1 to 0.5, are also tested, and the short timescale declining
trend is found to be more conspicuous with larger $\sigma_{norm}$.  These
results indicate that, should there be any covariance signal lurking in the
observed flux measurements, it would exert a declining trend, rather than a
rising trend, to the observed $s-\tau$ curves.  Again, despite impact of
measurement covariance on color variability calculation, it is not responsible
for the rising trend of the $s-\tau$ curves.

To conclude, in NGC 5548, the observed rising trend on the timescale dependency
of color variability is not any result of observational effects, and the color
variability predicted by the reprocessing scenario cannot match the timescale
dependent color variability trend we observed.

\section{Discussion}\label{sect:diss} 
\subsection{Known challenges to the reprocessing diagram}\label{sect:disk1} 
The reprocessing model was first brought up by \cite{1988MNRAS.233..475G}.
Though it usually neglects the origin of the intrinsic variability of the driven
source, it has the merit of offering an elegant explanation for the variability
in UV/optical bands and even in near infrared
\citep{2010IAUS..267...90L,2011MNRAS.415.1290L,2016ApJ...821...56F}.  Within the
simple reprocessing diagram, the in-phase variability and the inter-band lags
can both be nicely interpreted.  It is thus widely accepted in literature though
it has been constantly invoked and questioned over the past decades.
Furthermore, it may also hold the possibility of making AGNs the ``standard
candle'', as demonstrated in \cite{2007MNRAS.380..669C}.

Generally, the central illuminating source is assumed to be X-ray emission from
the corona, as a natural physical extension of the prevailing disk-corona model.
This is supported by numerous observations showing X-ray variations usually lead
UV/optical variations
\citep[e.g.,][]{2009MNRAS.399..750B,2010MNRAS.403..605B,2012MNRAS.422..902C,2014ApJ...788...10L,2014ApJ...788...48S,2014MNRAS.444.1469M,2015ApJ...806..129E,2017MNRAS.466.1777P}.
Indeed, recent measurements based on both X-ray reverberation mapping and
gravitational microlensing suggest that the corona is quite compact and located
in the vicinity (less than 20 $Rg$) of the SMBH
\citep{2012ApJ...756...52M,2013ApJ...769L...7R,2013ApJ...769...53M,2014MNRAS.438.2980C,2015ApJ...798...95B},
and thus likely to be a natural illuminating source.  However, there are still a
number of observations showing no clear correlation between X-ray and UV/optical
variations \citep[e.g.,][]{2002AJ....124.1988M,2006ASPC..360..111G}.
Specifically, for NGC 5548, based on the reprocessing diagram,
\cite{2017ApJ...835...65S} derived the driving light curve for the UV/optical
variations, and found it is poorly correlated with the observed \Swift{} X-ray
light curves.  Also, as was pointed out by \citet{2017MNRAS.470.3591G}, disk
reprocessing of the hard X-ray light curve produces UV/optical light curves with
too much fast variability, comparing with the observed ones.  They instead
proposed that what is reprocessed is not the hard X-ray emission, but probably
soft X-ray/FUV emission from the inner accretion disk with much smoother
variation.

The energy budget is another prominent problem if X-ray corona is the driving
source in the reprocessing diagram, as the luminosity of X-ray usually makes up
only a small portion of the total bolometric luminosity; hence it would be
insufficient to drive the variation of the whole disk
\citep{2007arXiv0711.1025G,2007ASPC..373..596G}.  This issue would be especially
serious for brighter quasars as the portion of X-ray luminosity is even smaller
\citep{2010A&A...512A..34L,2005AJ....130..387S,2010ApJS..187...64G}.

On more issue with the illumination scenario, is that strong Lyman continuum
emission would be expected on the spectra \citep{1997ApJ...476..605S,
1989ApJ...342...64A}. However, it has never been observed for AGNs.

The more recent issue revealed for the reprocessing diagram is that the observed
lags observed in many sources are systematically larger than predicted by the
standard thin disk theory, such as MR 2251-178 \citep{2008MNRAS.389.1479A}, PG
1211+143 \citep{2009MNRAS.399..750B}, AGN 0957+561 \citep{2012ApJ...744...47G},
NGC 3516 \citep{2016ApJ...828...78N}, Fairall 9 \citep{2017MNRAS.466.1777P} and
Pan-STARRS quasar sample \citep{2017ApJ...836..186J}.  Typically, it is found
that the observed lags are two to three times the lags given by the thin disk
model.  Efforts have been made in literatures to tackle this problem.  For
example, \cite{2017MNRAS.470.3591G} tried to reconcile this issue by pushing the
reprocessing region outward to the broad line region \citep[also see
][]{2018arXiv180306090S}. And as demonstrated in \cite{2016ApJ...828...78N}, the
accretion flow may form a radiatively inefficient accretion flow, affecting the
reprocessing paradigm, as well as the time delays, but this scenario is limited
to accretion systems with low Eddington ratio. 

\subsection{A new evidence against the reprocessing diagram} In this work, we
find the UV/optical color variation in NGC 5548 is clearly timescale-dependent,
a trend similar to that found in quasars, but for the first time detected in an
individual AGN. 

We present a simple reprocessing model showing that reprocessing model seems to
able to well reproduce the observed UV/optical light curves and the inter-band
lags (see Section~\ref{sect:disk1} for caveats of the reprocessing model).
However, severe inconsistency between the reprocessing model and observations
emerges while we look into the timescale dependency of the color variation.  The
reprocessing model failed to recover the observed timescale-dependent pattern in
color variation.  This can be further explained as below.  In the reprocessing
scheme, the reprocessed light curves are smoothed and lagged version of that of
the driving source.  The lag also corresponds to the smoothing timescale. At
timescales considerably longer than the lags, reprocessed light curves in
different bands are expected to be rather similar, thus is unable to reproduce a
timescale-dependent color variation pattern.  Unlike other challenges to the
reprocessing diagram we discussed in Section~\ref{sect:disk1}, such controversy
is hard to be reconciled within the reprocessing scheme.  Therefore,
reprocessing should not be the dominant mechanism in producing the observed
UV/optical variation in NGC 5548.  This also demonstrates that the timescale
dependency of the color variation is a unique and powerful probe in diagnosing
the AGN variations. 

It is worthwhile to mention that the accretion disk size in high redshift
quasars measured through micro-lensing studies is about 4 times that of the
standard thin disk model
\citep{2007ApJ...661...19P,2010ApJ...718.1079B,2010ApJ...712.1129M,2010ApJ...709..278D}.
By splitting the accretion disk into large number of individual zones, and
allowing the temperature of each zone fluctuating significantly and
independently, \cite{2011ApJ...727L..24D} found an inhomogeneous disk model is
able to resolve such discrepancy in the disk size.  More interestingly, a
revised version of the inhomogeneous disk model is able to naturally yield a
timescale-dependent color variation, consistent with observations
\citep{2016ApJ...826....7C}.
 
However, such inhomogeneous disk model is not yet perfect. In the model, the
temperature of each region of the disk varies independently, leading to no
correlation between the variation of each individual disk region, thus faces
difficulty to explain the observed well coordinated multi band light curves
\citep{2015MNRAS.449...94K,2016ApJ...826....7C}.  Furthermore, the current model
is unable to explain the observed inter-band lags.  A further revised
inhomogeneous disk model appears able to reproduce the observed inter-band lags
and coordination without the need of light echoing \citep{2018ApJ...855..117C}.

Alternatively, both X-ray reprocessing and disk fluctuations might be at work,
while the dominance of each might be epoch and timescale-dependent, and vary
from source to source.  Note some researches
\citep{2008MNRAS.389.1479A,2009MNRAS.394..427B} have shown that the X-ray
reprocessing would work on shorter timescales but fail to explain the
variability on longer timescales.  Also, \cite{2009MNRAS.397.2004A} noticed that
the variability amplitude in optical bands on longer timescale was greater than
in X-ray in NGC 3783, thus can not be attributed to X-ray reprocessing. It is
likely that reprocessing might play a major role at very short timescales, while
disk fluctuation dominate at longer timescales.  A dedicated quantitative work
is also ongoing to investigate whether such a hybrid model is able to explain
all the observations, in terms of inter-band correlations and lags, energy
budget, disk size, and timescale-dependent color variation.

To summarize, we show that the color variation in NGC 5548 is clearly
timescale-dependent, significantly challenging the widely accepted reprocessing
diagram.  A revised disk fluctuation model without the need of light echoing
\citep{2018ApJ...855..117C} seems able to well reproduce the observed variation
in NGC 5548, including inter-band coordination and lags, and timescale-dependent
color variation.  Extensive work is also ongoing to investigate a hybrid model
including both X-ray reprocessing and disk fluctuations.  In addition, we
demonstrate the power of timescale dependency of color variability again.  It is
possible to look into the inner part of the accretion disk with high cadence
light curves.  The idea of exchanging for physical resolution using temporal
resolution is worthy exploring further in the age of temporal astronomy.

\section*{Acknowledgment} We are grateful for the anonymous referee's careful
reading and insightful reports, which improve this work a great deal.  We also
thank M. Mehdipour for kindly providing us their \Swift{} light curves, which
were used in our early analyses.  Special thanks go to Robert Antonucci for
bringing the Lyman continuum issue for illumination scenario to our attention.
This work is supported by National Basic Research Program of China (973 program,
grant No. 2015CB857005) and National Science Foundation of China (grants No.
11233002, 11421303 $\&$ 11503024).  J.X.W. thanks support from Chinese Top-notch
Young Talents Program, and  CAS Frontier Science Key Research Program
QYCDJ-SSW-SLH006.  Z.Y.C. acknowledges support from the Fundamental Research
Funds for the Central Universities.

\software{Astropy \citep{2013A&A...558A..33A}, Matplotlib \citep{Hunter:2007ih},
JAVELIN\citep{2011ApJ...735...80Z,2013ApJ...765..106Z}, PYCCF \citep{Sun2018},
sour\citep{2017ApJ...840...41E}}.

\end{CJK*} \bibliography{ref} \end{document}